\documentclass[journal,comsoc]{IEEEtran}
\usepackage{subcaption}
\usepackage{caption}
\usepackage{comment}
\usepackage{subfig}
\usepackage{tikz}
\usepackage{graphicx}
\usepackage{enumerate}
\usepackage{amsmath,amssymb,amsfonts}
\usepackage{algorithmic}
\usepackage{lscape}
\usepackage{mathptmx}
\usepackage{float}
\usepackage{array}
\usepackage{hhline}
\usepackage{url,multirow,multicol}
\usepackage{color,soul}
\usepackage{pifont}
\usepackage[export]{adjustbox}
\usepackage[ruled,vlined,linesnumbered]{algorithm2e}
\usepackage[nointegrals]{wasysym}
\usepackage{textcomp}
\usepackage{amssymb}
\usepackage{blindtext}
\SetKwInput{KwInput}{Input}
\SetKwInput{KwOutput}{Output}

\usetikzlibrary{arrows}  
\usepackage{pgfplots}
\pgfplotsset{compat=1.12}

\graphicspath{{images/}} 

\ifCLASSINFOpdf

\else
  
\fi

\usepackage{amsmath}

\usepackage[cmintegrals]{newtxmath}
\graphicspath{{pics/}}

\begin{document}

\twocolumn

\title{Novel Rewiring Mechanism for Restoration of the Fragmented Social Networks after Attacks}

\author{Rajesh~Kumar,
        Suchi~Kumari,
        Anubhav Mishra

\thanks{R. Kumar is with Amity School of Engineering and Technology, Amity University, Bangalore, India. (Email: rkumar1@blr.amity.edu).

S. Kumari is with the Department of Computer Science and Engineering, Shiv Nadar Institute of Eminence, Delhi-NCR, India (Email: suchi.singh24@gmail.com).

A. Mishra is with the Department of Electrical Engineering, Ghani Khan Institute of Engineering and Technology, Narayanpur Malda, India (Email: tooli.anubhav@gmail.com).

}
}


\maketitle
\begin{abstract}
Real-world complex systems exhibit intricate interconnections and dependencies, especially social networks, technological infrastructures, and communication networks. These networks are prone to disconnection due to random failures or external attacks on their components. Therefore, managing the security and resilience of such networks is a prime concern, particularly at the time of disaster. Therefore, in this research work, network is reconstructed by rewiring/addition of the edges and robustness of the networks is measured. To this aim, two approaches namely (i) Strategic rewiring (ii) budget constrained optimal rewiring are adopted. While current research often assesses robustness by examining the size of the largest connected component, this approach fails to capture the complete spectrum of vulnerability. The failure of a small number of connections leads to a sparser network yet connected network. Thus, the present research work delves deeper into evaluating the robustness of the restored network by evaluating Laplacian Energy to better comprehend the system's behavior during the restoration of the network still considering the size of the largest connected component attacks.

\end{abstract}


%
\IEEEpeerreviewmaketitle

\section{Introduction}
\IEEEPARstart{R}{eal}-world networks are susceptible to failure due to the malfunction of certain components or deliberate strategies. Cascading failures and targeted attacks can severely disrupt network operations, affecting areas such network controllability \cite{nie2014robustness}, the dynamic process, such as the spreading of diseases \cite{moreno2004dynamics}, synchronization \cite{kumar2024effect}, connectivity \cite{kumar2021minimizing}, and more. To ensure the smooth operation of such networks, it is crucial for nodes and links to be resilient against both random and targeted attacks, especially during disasters. This resilience can be achieved by rewiring existing edges or adding new connections. Traditional methods assess robustness primarily by measuring the size of the largest connected component \cite{kumari2024novel}, but this approach fails to capture all vulnerabilities, as minor failures may still leave the network connected. Hence, this work offers a more comprehensive evaluation of network robustness by using Laplacian Energy, which provides a deeper understanding of the network's behavior during restoration, while still considering the size of the largest connected component after attacks.

The complex network properties, such as centrality measures, giant connected components, entropy \cite{sridhar2022entropy}, can be used for performance evaluation of the network under various circumstances, e.g., information spreading in social networks \cite{wan2021constrained}, network attack, random failure \cite{kumar2021minimizing}, network congestion \cite{kumari2020efficient1}, efficient network topology design, etc.  Holme \textit{et al.} \cite{holme2002attack} studied the performance of complex network models and some real-world networks under attacks on nodes and links. The critical point of failure is evaluated using Monte Carlo simulations for the networks with scale-free and bimodal degree distributions \cite{paul2005resilience}. Zhang \textit{et al.} \cite{zhang2015notion} have used the concept of r-robustness which guarantees that the network will remain connected even if a certain number of nodes are removed from the neighbourhood of every node in the network.  Kumari et al. \cite{kumari2020efficient1} introduced link rewiring strategies to enhance the transport efficiency of scale-free networks. In these networks, new nodes often connect to hub nodes, which can become congested due to the high traffic flow passing through them. The proposed rewiring strategy aims to reduce the centrality of high-betweenness nodes while maintaining the network's degree distribution.  

Kumar \textit{et al.} \cite{kumar2021privacy} proposed a Privacy-Preserving Rewiring Algorithm (PPRA) that uses fuzzy logic to anonymize social network data, and the approach is validated through real-world datasets and demonstrates effectiveness across three major graph mining tasks.  Farine \cite{farine2021structural} introduced a second-degree rewiring strategy for maintaining network structure after node failure. This approach encourages nodes that lose a shared connection to form new links with each other, helping preserve clustering, density, and the clustering coefficient with less disruption than random rewiring. Kumar \textit{et al.} \cite{kumar2023robustness} evaluated the robustness of the multilayer network considering the graph energy concept. Keenland \textit{et al.} \cite{kneeland2024rewiring} studied the effects of corporate offsite visits on employee connections, showing that participants initiate and receive more collaborative ties, ultimately enhancing collaboration and benefiting both individuals and the organization.

Several research studies have examined the role of community structure in assessing network robustness. Ma \textit{et al.} \cite{ma2013enhancing} considered groups of nodes, or communities, to evaluate the resilience of networks. They addressed a two-level network attack scenario and assessed how the integrity of the social community structure influences information fusion within the network. Kumari \textit{et al.} \cite{kumari2021intelligent} introduced intelligent community deception techniques designed to hide nodes from community detection algorithms, aiming to protect privacy in industrial networks. They also employed rewiring strategies to reduce persistence scores and enhance safety scores. Borges \textit{et al.} \cite{borges2024social} examined how homophilic and heterophilic rewiring affect consensus formation in social networks. They found that in complex opinion diffusion, heterophilic rewiring helps bridge polarized communities, while homophilic rewiring promotes consensus in simple diffusion. A combination of both rewiring preferences reduces polarization and fosters consensus across different diffusion processes. 

Several research works have explored neuron rewiring in artificial neural networks. Landmann \textit{et al.} \cite{landmann2021self} proposed a local rewiring strategy based on node activity, leading to a self-organized network that reaches a critical state and exhibits power-law avalanches. Li \textit{et al.} \cite{li2024adaptive} introduced adaptive rewiring for brain networks, reorganizing them based on internal signal intensity to create modular, small-world structures. Scabini \textit{et al.} \cite{scabini2024improving} focused on neuronal centrality and strength variance in ANNs, presenting a preferential attachment rewiring method to optimize weight distribution for better deep ANN initialization.

From the literature survey, it can be observed that the failure of complex systems is inevitable. Some studies focus on adding new connections to improve network robustness, while others propose rewiring strategies, which are generally more cost-effective. Although simple topology alterations can enhance robustness under attack \cite{louzada2013smart, rebouccas2019energy, kumari2021intelligent}, random rewiring increases more costs in terms of effort, money, etc. Since each rewiring operation incurs a cost, random rewiring leads to higher expenses compared to strategic approaches \cite{xia2021graph}. In most of the research works, robustness is often assessed by the size of the largest connected component (LCC), this method fails to fully capture vulnerability, as small failures can still leave a network connected. Therefore, rewiring should be both strategic and budget-constrained, with algorithms learning from prior operations to ensure connectivity with minimal cost. The total cost must remain within the available budget, necessitating a bounded-constrained framework to enhance robustness and manage node failures.

In the present work, we not only propose the different rewiring methods but also, the cost involved in the rewiring strategies is also analyzed. In many of the research works cited above, the robustness of the network is measured by the size of the largest connected component (LCC) where it provides only the fraction of nodes present LCC irrespective of the topology of the underlying network e.g., connectivity patterns. Now, in a tree network, LCC can have $n-1$ edges and by removing only a fraction of edges, the LCC may result in multiple disconnected component. Whereas, knowing the connectivity patterns (graph density, node degree etc.) along with the number of nodes in LCC, one can estimate the robustness of the network. Therefore, in the present research, connectivity patterns are considered based on which, Laplacian network energy is evaluated in networks with different topological structures.

By considering all the discussed issues, the key offerings of the manuscript are as follows:
\begin{enumerate}[1.]
    \item A framework is provided to determine the edge weights based on the trust score of the nodes. The edge weights are used to quantify the importance of the edges and hence are proportional to the cost of the edge.
    \item Rewiring approaches (Strategic and budget-constrained) proposed to restore the network.
    \item Robustness of the restored of network is evaluated by computing the Laplacian energy of the graph and robustness index.
\end{enumerate} 

The rest of the paper is organized as follows. Section II defines various network parameters. Section III discusses the proposed attack strategies and various rewiring approaches. The results and detailed analysis are provided in section IV. Section V concludes the proposed research work and provides future directions.

\section{Proposed Methodology}\label{method}

The real-world network is prone to attack, and in the worst case, it breaks the whole network into multiple components. The network component, such as nodes or edges may be vulnerable, so it is required to restore the entire network up to some desirable levels. There may be multiple ways to restore the network e.g., repairing of damaged links, addition of redundant links \cite{chaoqi2018complex}. In the present research work, a novel method of restoration of the network is proposed where the rewiring (of existing links) / addition of new links mechanism is adopted under budget constraints. For the simulation purpose, initially, we consider a complete network of a given size and edge weights are calculated according to the method provided in the subsection below. 

\subsection{Edge weight computation}\label{edge_weight}
Edge weights of a network can be conceptually interpreted in many different ways. For instance, tracking individual interactions over a predetermined amount of time in social networks might provide a representative weight or indicator of link strength. In order to calculate edge weights, the node's trust value and the degree of connectivity between nodes are considered. 
\subsubsection{Calculation of trust value of the nodes}\label{trust-val} 

The development of trust connections and the dependability of interactions among network nodes are crucial for a variety of networks, including social, financial, and communication networks. Without a strong analytical and conceptual framework, it might be difficult to understand and evaluate trust. In this part, we suggest a mathematical and statistical methodology for assessing the dependability of network nodes. We apply a well-known statistical method called the Maximum Likelihood Estimation method to measure the overall trust value ($\phi$) associated with a particular node $i$ within a multi-layer network system. As described in other studies \cite{gatti2004probability,szekely1986paradoxes}, this strategy entails adapting a mathematical model to the available data. Let $n$ be the total number of nodes and $\mathcal{S}$ and $\mathcal{F}$ signify matrices of dimensions ($n \times n$). The numbers of successful and failed transactions between nodes $i$ and $j$ are recorded in these matrices. The following formulation is offered.
\[
    s_{ij}= 
\begin{cases}
    T_{ij} & \text{if there is an edge between} \hspace{2mm} i \hspace{2mm}\text{and}\hspace{2mm} j \\
    0,              & \text{otherwise} 
\end{cases}
\]

\[
    f_{ij}= 
\begin{cases}
    U_{ij} & \text{if there is an edge between} \hspace{2mm}i \hspace{2mm} \text{and} \hspace{2mm}j\\
    0,              & \text{otherwise} 
\end{cases}
\]
where  $(T_{ij})$ and  $(U_{ij})$ are a number of successful and failed transactions. We define the terms $\psi_{i}=\displaystyle \sum_{i=1,i \neq j}^{n}s_{ij}$ and $\eta_{i}=\displaystyle \sum_{i=1,i \neq j}^{n}f_{ij}$ to represent the aggregate of successful and failed transaction such that likelihood function $\phi_{i}$ is provided in Eq. \eqref{eq1}.
\begin{equation}
    L(\phi_{i},y_{i,1},y_{i,2},\dots,y_{i,n})=\displaystyle \prod_{j=1,j \neq i}^{n} ({\phi_{i}})^{\psi_{i}} (1-\phi_{i})^{\eta_{i}}
    \label{eq1}
\end{equation}
Log-likelihood function ($\zeta$) of $\phi_{i}$ from Eq. \eqref{eq1} can be written as,
\begin{equation}
    \zeta(\phi_{i})=\text{ln} \hspace{1mm}L(\phi_{i})=\psi_{i} \text{ln}\hspace{1mm}\phi_{i} + \eta_{i} \text{ln}\hspace{1mm} (1-\phi_{i})
    \label{eq2}
\end{equation}
By differentiating Eq. \eqref{eq2}, likelihood equation is obtained as follows,
\begin{equation}
    \frac{d\zeta(\phi_{i})}{d\phi_{i}}=\frac{\psi_{i}}{\phi_{i}}-\frac{\eta_{i}}{1-\phi_{i}}=0
    \label{eq3}
\end{equation}
Thereafter, Eq. \eqref{eq3} is used to determine the MLE of $\phi_{i}$, i.e., the probability of the trust value of any node $i$ to provide quality services. 
 
 \subsubsection{Calculation of edge weights} \label{edge_weights}
 
 The link weights ($w_{i,j}$)
are determined based on the trust values ($\phi)$ and the connectivity pattern ($a_{ij}$) of the nodes. Therefore, the link weights can be expressed as follows.

\begin{equation}
    w_{ij}=\mathcal{F}(\Gamma,a_{ij})=\Gamma+a_{ij}
    \end{equation}
    \begin{equation}
        \Gamma(\phi_{i},\phi_{j})=\Big{[\Big(} \frac{\text{Cos}(\Delta_{ij})+1}{2}\Big{)}\times(\phi_{i}\times\phi_{j})\Big{]}
    \end{equation}
where $\Delta_{ij}=|\phi_{i}-\phi_{j}|$. The rationale behind employing the Cosine function is as follows. The Cosine function's range, denoted as Cos($\omega$), spans from $-1$ to $1$ for any real value $\omega \in \mathbb{R}$. Consequently, to normalize the Cos($\omega$) values within the range of $0$ to $1$, the transformation $\frac{Cos(\omega)+1}{2}$ is utilized. This normalization process facilitates the conversion of the cosine values into a scale that falls between $0$ and $1$.
Now, consider two distinct scenarios:
\begin{enumerate}[(i)]
    \item  When $\phi_{i}=0.1$ and $\phi_{j}^{\alpha}=0.1$.
    \item When $\phi_{i}=0.99$ and $\phi_{j}^{\alpha}=0.99$.
\end{enumerate}

In both cases, the computed trust difference $\Delta_{ij}^{\alpha}=0$, which leads to the value of $\frac{Cos(\omega)+1}{2}$ being equal to $1$. However, an interesting observation arises: the trust values in scenario (ii) are considerably higher than in scenario (i), despite both resulting in the same $\frac{Cos(\omega)+1}{2}$ value. To address this discrepancy and appropriately scale the trust values, a scaling factor is introduced. This scaling factor is calculated as the product of the trust values for nodes $i$ and $j$, i.e., ($\phi_{i}^{\alpha}\times\phi_{j}^{\alpha}$). This factor is then multiplied by $\frac{Cos(\Delta_{ij}^{\alpha})+1}{2}$, ensuring that the trust values are adjusted in a manner that accurately reflects their magnitudes.

\subsubsection{Robustness Evaluation using Graph Laplacian Energy Concept}

In this section, a mathematical framework is presented for assessing the robustness of  network systems through the utilization of the graph's Laplacian energy. In previous research works, scholars have embraced the use of the Normalized Laplacian matrix to ascertain the robustness of networks \cite{gao2013r}. However, since the Normalized Laplacian matrix doesn't provide insights into the network's edges, an alternative approach is needed. Thus, to compute the robustness of the considered networks, the Laplacian energy framework is adopted. In this context, robustness is defined in terms of the Laplacian energy denoted as $L_{\mathcal{E}}$. This Laplacian energy quantifies the robustness of the entire network, and its formulation can be expressed as follows.

\begin{equation}
   L_{\mathcal{E}}=\frac{1}{n}\sum_{i=1}^{n}(\lambda_{i}-{\Bar{\lambda}})^2
\label{l_energy_1}
\end{equation}

where $\lambda_{i}$ is the $i^{th}$ eigenvalue of Laplacian matrix $\mathcal{L}$ and ${\Bar{\lambda}}=\frac{1}{n} \displaystyle\sum_{i=1}^{n}\lambda_{i}$. The naive approach to calculate $L_{\mathcal{E}}$ is computationally expensive. Therefore, we analyze the spectrum of the Laplacian matrix and propose a simple and efficient approach to compute $L_{\mathcal{E}}$. The Laplacian graph eigenvalues for the networks obey the following relations \cite{gutman2006laplacian},
\begin{equation}
    \sum_{i=1}^{n}\lambda_{i}=2m
    \label{l_energy_2}
\end{equation}
\begin{equation}
    \sum_{i=1}^{}(\lambda_{i})^2=2m + \sum_{i=1}^{n}(k_{i})^2
    \label{l_energy_3}
\end{equation}

where $m$ is the total number of edges of graph and $(k_{i})$ is the degree of node $i$. Now, using Eq. \eqref{l_energy_1}, $L_{\mathcal{E}}$ is derived as follows,

\begin{equation}
  L_{\mathcal{E}}=  \frac{1}{n} \Big[\sum_{i=1}^{n}(\lambda_{i})^2 + n({\Bar{\lambda}})^2- 2\times{\Bar{\lambda}}\times\sum_{i=1}^{n}{\lambda_{i}}\Big]
  \label{l_energy_4}
\end{equation}
Now, from Eqs. \eqref{l_energy_2},\eqref{l_energy_3},\eqref{l_energy_4} and ${\Bar{\lambda}}$, we have,
\begin{equation}
    L_{\mathcal{E}}=\frac{1}{n} \Big [2m + \sum_{i=1}^{n}(k_{i})^2 -\frac{4 \times m^2}{n}\Big]
    \label{l_energy_6}
\end{equation}
Eq. \eqref{l_energy_1} makes it easier to carry out the previous derivation, and more crucially, it does so without the need to compute the eigenvalues of the Laplacian matrix. Eq. \eqref{l_energy_2} provides a link between the total number of edges in the network and the sum of the eigenvalues of the Laplacian matrix. Similar to Eq. \eqref{l_energy_2}, Eq. \eqref{l_energy_3} describes the connection between the summation of the squared eigenvalues of the Laplacian matrix, the total edge count, and the summation of the squared node degrees. The mathematical equation for the Laplacian energy, denoted as $L_{\mathcal{E}}$, is derived from Eq. \eqref{l_energy_2} and Eq. \eqref{l_energy_3} as shown in Eq. \eqref{l_energy_6}.

\subsubsection{Robustness index ($S$)} The component in a graph having a maximal connected subgraph is called a giant component. In other words, when the size of the connected component (number of nodes) $\rightarrow N$ (total number of nodes) then the connected component is called a giant component. The normalized Giant$/$Largest Connected Components (LCC) can be calculated using Eq. \eqref{eqn:LCC}.
\begin{equation}
    S = \frac{Size of LCC}{n} \label{eqn:LCC}
\end{equation}
where LCC is the largest connected component and $n$ is the total number of nodes in the network.

\subsubsection{Graph density ($\rho$)} It is the ratio of the number of edges of a graph to the maximum number of edges that a graph can have. It indicates the degree to which an edge in a graph is conceptually tightly connected. When we want to expand an extended network by adding additional edges, it's quite useful. Graph density also aids in our estimation of the remaining number of edges that need to be added to the network. Mathematically, graph density ($\rho$) for an undirected graph is given in Eq. \eqref{rho}.
\begin{equation}
    \rho=\frac{2m}{n(n-1)} \label{rho}
\end{equation}
Relation between Laplacian energy $ L_{\mathcal{E}}$ and graph density ($\rho$) is given by,

\begin{equation}
    L_{\mathcal{E}}=\rho(n-1)[1-\rho(n-1)] + \frac{1}{n}  \sum_{i=1}^{n}(k_{i})^2
    \label{l_energy_7}
\end{equation}
Note: For $d$ regular graph, $L_{\mathcal{E}}=d$.

\subsection{Rewiring Mechanism} \label{rewire_mech}
In the rewiring process, depending upon the number of edges of the largest connected component, either edges are rewired or new edges are introduced. For example, for any network to be connected, it requires minimum ($n-1$) edges where $n$ is the number of nodes. There may be a case where the largest connected component does not contain ($n-1$) edges therefore, some extra connections are to be introduced otherwise, rewiring will lead to the formation of a new disconnected component and we will never get the restored network.
In the present work, two mechanisms ate adopted for rewiring (i) Strategic and (ii) Based on budget constraints. These are explained in the following sections.

\subsubsection{Strategic rewiring}
In this approach, a node $i$ with the maximum degree is selected from the Largest Connected Component (LCC), and one of its neighbors $j$ with $k_j \geq 2$ is chosen for rewiring. If there are multiple neighboring nodes with the same degree, one is selected randomly. After selecting the appropriate neighbor $j$ for rewiring, the edge $e_{ij}$ is disconnected from the node $j$ and reconnected to a node $l$ with the maximum degree from the disconnected component that contains more nodes. This strategy is based on the observation that the energy of the network, as given by Eq. \eqref{l_energy_6}, mainly depends on the degrees of the nodes. Since any node from the disconnected component to the LCC will result in the same number of edges and nodes in the newly connected component, the degree of the nodes is the only factor influencing the network's energy. If the number of edges is less than $n-1$, a new connection is made between node $i$ from the LCC and node $l$ (with the maximum degree, as mentioned in the rewiring procedure), and the rewiring continues until all disconnected components are reconnected. Followings are obtained after each rewire operation
(i) ${{L}^{r}_\mathcal{E}}$; the Laplacian energy of the newly formed LCC after $r^{th}$ rewiring (ii) $\mathcal{C}^{r}$;cost of $r^{th}$ rewired/added link
(iii) ${i}^{r}$; $i^{th}$ node picked from LCC for $r^{th}$ rewire/addition of link (iv) ${j}^{r}$; $j^{th}$ node picked from one of the disconnected components for $r^{th}$ rewire/addition of link. The process rewire$/$addition of the links is represented in Algorithm \ref{algo3}. 

\begin{figure}[!htb]
\begin{center}
$\begin{array}{ccc}
\includegraphics[width=.5\linewidth,height=1.5in]{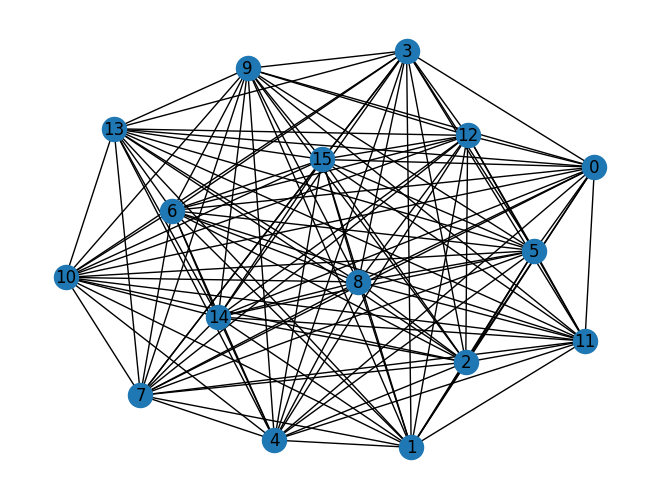}  &
\includegraphics[width=.5\linewidth,height=1.5in]{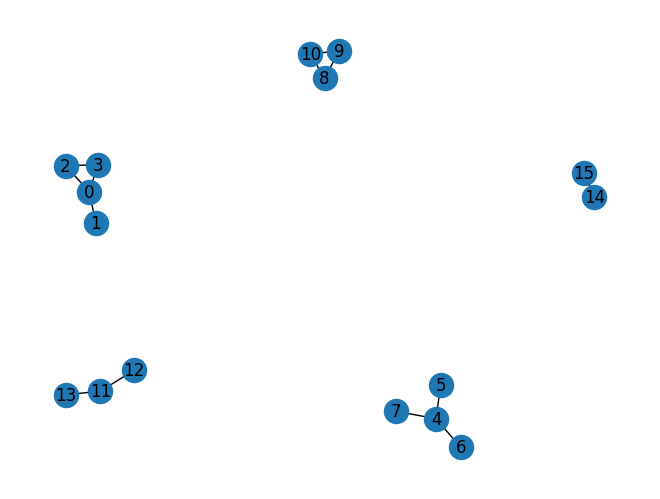} \\
\mbox{(a)} & \mbox{(b)} \\
\includegraphics[width=.5\linewidth,height=1.5in]{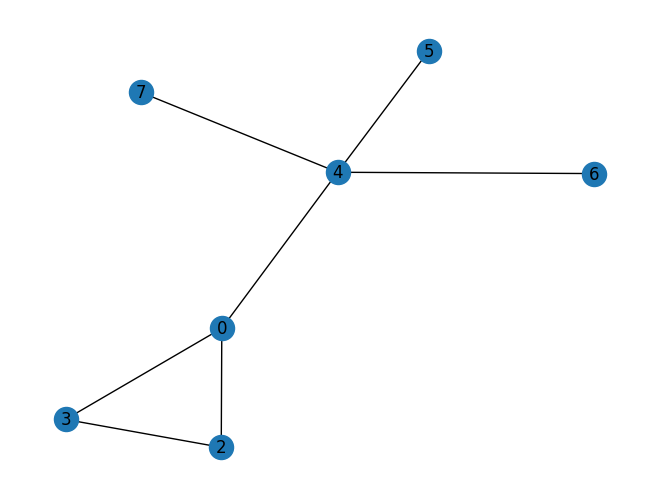}  &
\includegraphics[width=.5\linewidth,height=1.5in]{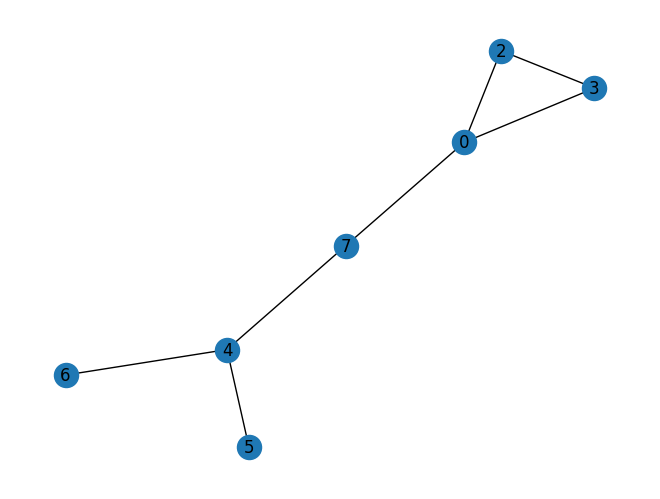} 
\\
\mbox{(c)} & \mbox{(d)} \\
\end{array}$
\caption{A toy network to illustrate the rewiring process.}
\label{Rewire_example}
\end{center}
\end{figure}

\begin{figure}[!htb]
\begin{center}
$\begin{array}{ccc}
\includegraphics[width=.45\linewidth,height=1.5in,frame]{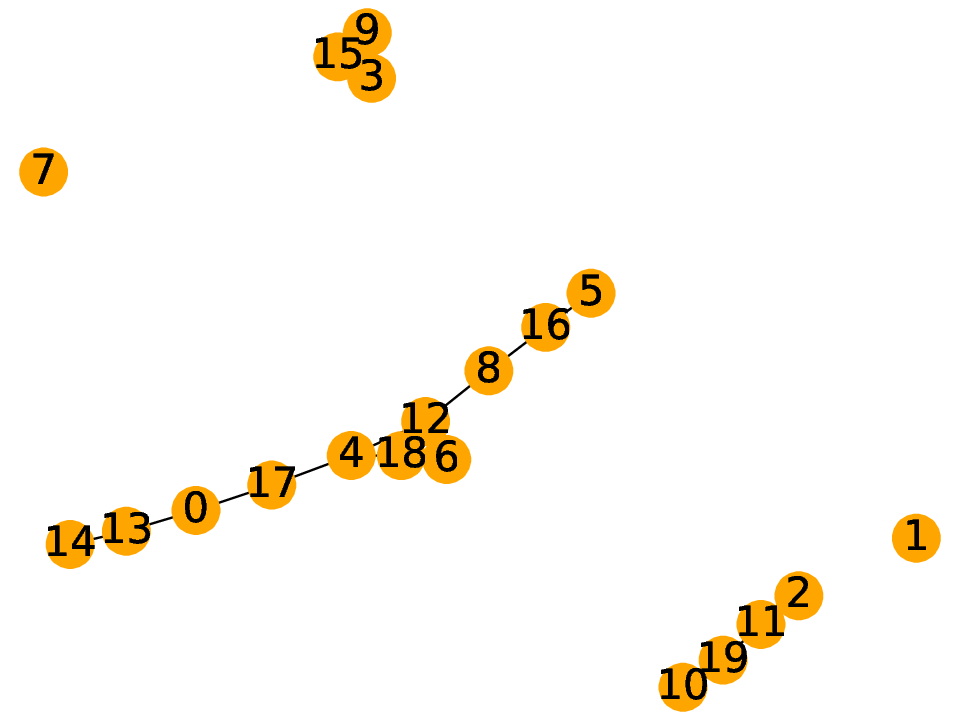}  &
\includegraphics[width=.45\linewidth,height=1.5in,frame]{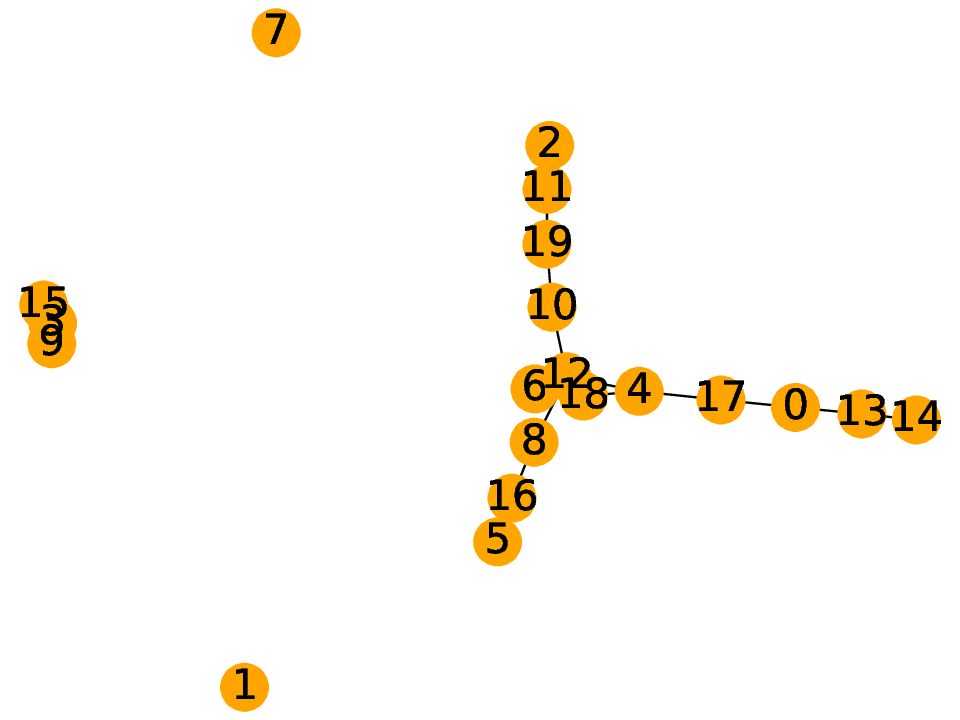} \\
\mbox{(a)} & \mbox{(b)} \\
\includegraphics[width=.45\linewidth,height=1.5in,frame]{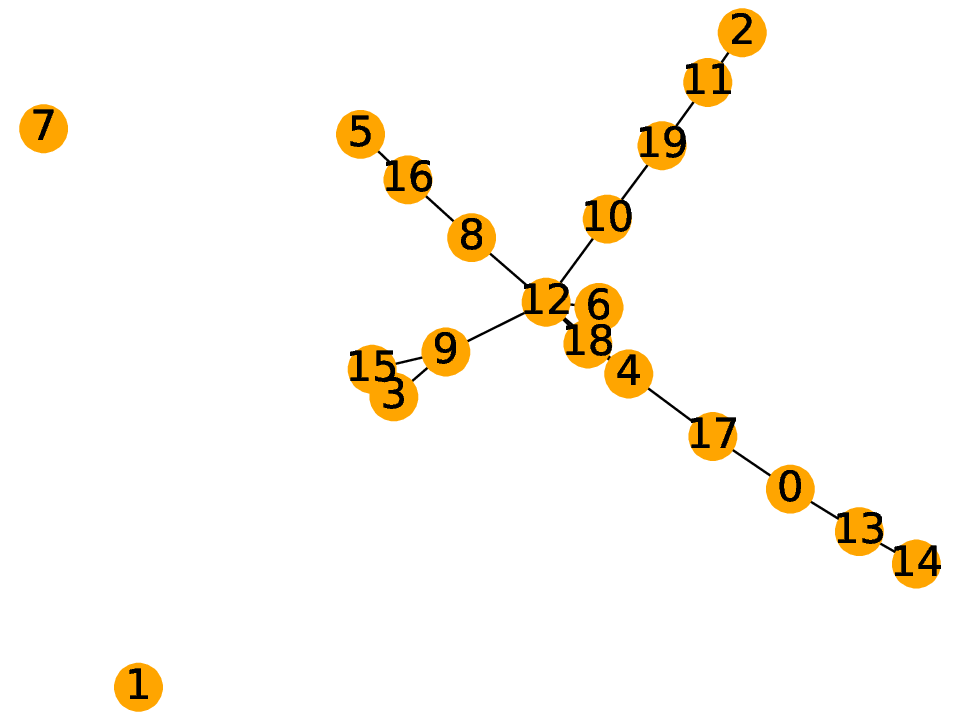}  &
\includegraphics[width=.45\linewidth,height=1.5in,frame]{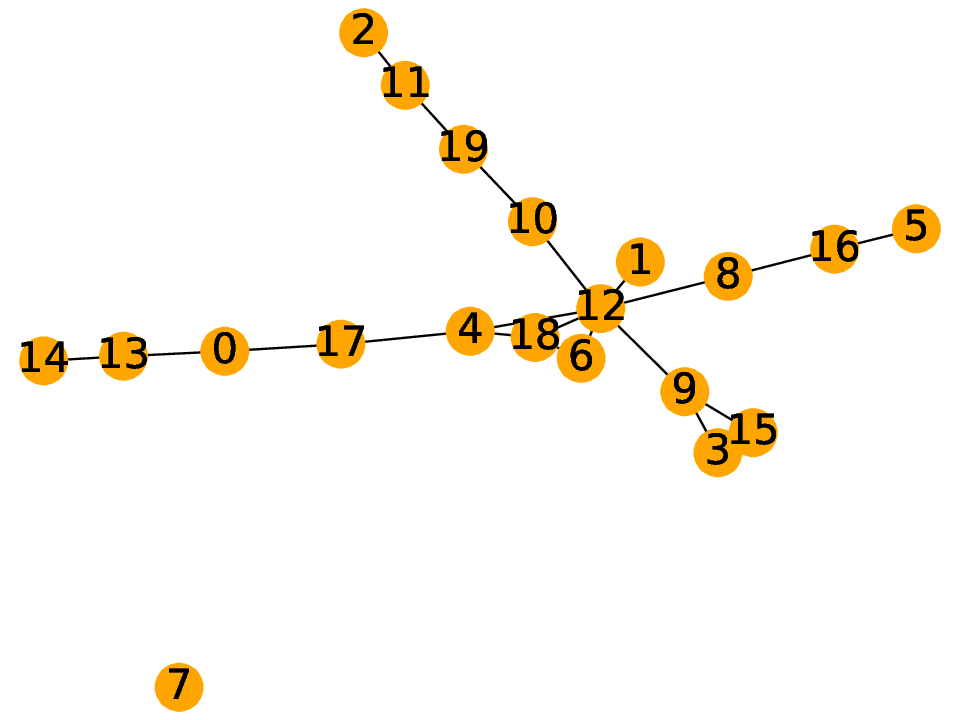} \\
\mbox{(c)} & \mbox{(d)}\\
\includegraphics[width=.45\linewidth,height=1.5in,frame]{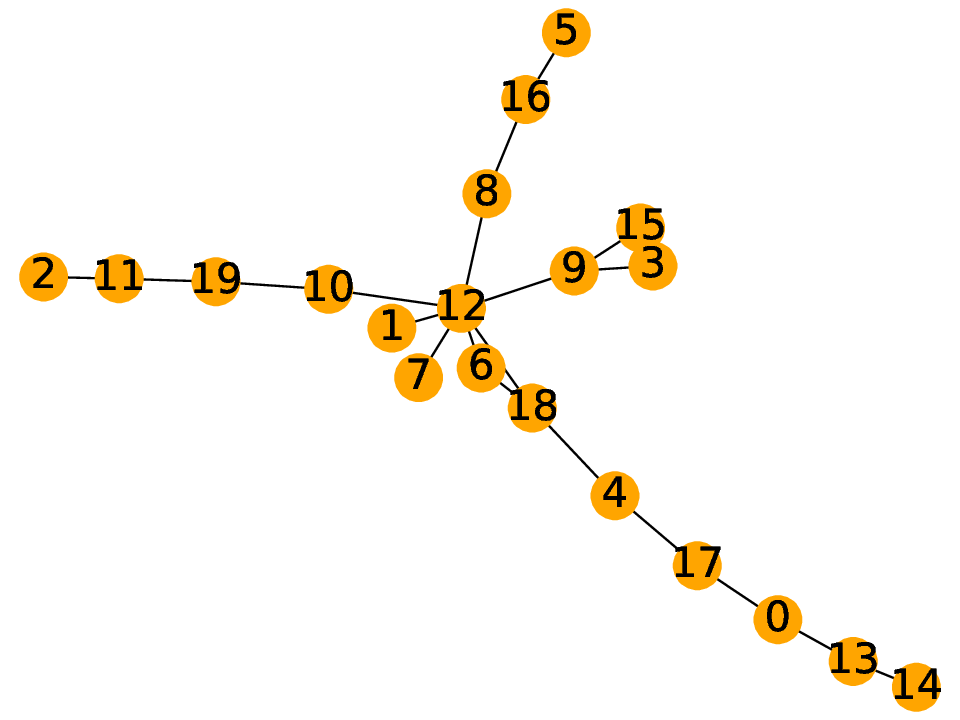}
\\
\mbox{(e)} \\
\end{array}$
\caption{(a) Initially, the network is under attack and splits into six components. The nodes in the Largest Connected Component (LCC) are $[0,4,5,6,8,12,13,14,16,17,18]$, while the remaining five disconnected components consist of the following node lists:$[[19,10,2,11],[9,3,15],[1],[7]]$. (b) Since the number of edges in the LCC is less than 20 (the number of nodes), an edge is added between node 12 (the node with the maximum degree in the LCC) and node 10 (the node with the maximum degree in the component $[19,10,2,11]$). (c) With the addition of the new edge in the LCC, the number of edges is still less than 20, so another edge is added, this time between node 12 of the LCC and node 9. (d) Once the number of edges in the LCC reaches 20, rewiring takes place according to Algorithm \ref{algo3}, and node 7 is connected to node 12 in the LCC. (e) The rewiring process continues, and eventually, the entire network is restored.}
\label{Rewire_example1}
\end{center}
\end{figure}

An example of the rewiring process is illustrated as follows: Initially, a complete network is considered with $n$ nodes (in Fig. \ref{Rewire_example}(a)), where in this case $n=16$, with node labels ranging from 0 to 15. After attacks (either random or strategic), edges are progressively removed from the network until it starts to disintegrate, reaching a threshold value for network energy. Beyond this threshold, edges continue to be removed until a predefined number of disconnected components are obtained (for example, $4$ components, each with $4$ nodes in the Largest Connected Component, or LCC). At this point, the rewiring process begins to restore the network, considering that each rewire has an associated cost and a limited available budget. As shown in Fig. \ref{Rewire_example}(b), assume that after the attacks, the LCC consists of four nodes (0-3). In the random rewiring approach, any edge connecting two nodes can be selected, with one end being disconnected and reconnected to a node from any of the disconnected components. For instance, in random rewiring, if an edge connecting nodes 0 and 1 is selected, it is disconnected from node 1 and reconnected to node 7. This rewiring results in 7 nodes in the LCC, as shown in Figure \ref{Rewire_example}(c), and the network energy $L_{\mathcal{E}} = 2.57$. In contrast, in the strategic rewiring approach, using the same edge selection, the other end is reconnected to node 4, yielding 7 nodes in the LCC as shown in Figure \ref{Rewire_example}(d), with $L_{\mathcal{E}} = 3.14$. Apart from that, a snapshot of simulations of a toy random network with $n=20$ and $p=0.5$ is shown in Fig. \ref{Rewire_example1}. In this case, the total number of links is less than $n-1$, so new links are added and connected to nodes with larger degrees. This process of adding new links continues until the network becomes fully connected.

\subsubsection{Rewiring based on budget constraint}
In this approach, rewiring is carried out in a two-fold approach: Firstly, we maintain four lists $\mathcal{E}$, $\mathcal{C}$, $\mathcal{I}$, $\mathcal{J}$ obtained from the strategic rewiring approach. Let, each element $\theta_r \in \mathcal{E}$ represents the energy of the LCC after $r^{th}$ rewiring and $c_r \in \mathcal{C}$ indicates the cost involved in $r^{th}$ rewiring, $\mathcal{I}$ is the list of candidate nodes from the LCC and $\mathcal{J}$ list of candidate nodes from the disconnected components. For the given budget $b$, the optimal set of links $\mathcal{E}^{\star}$ is obtained that maximizes the energy of the LCC as mentioned in Algorithm \ref{algo1}. In the second fold, budget $b$ is increased as 10 $\%$ of the total budget $B$ and Algorithm \ref{algo1} is called and $\mathcal{E}^{\star}$ is obtained. Corresponding to $\theta_r \in \mathcal{E}^{\star}$, $i_r \in \mathcal{I}$ and $j_r \in \mathcal{J}$ are chosen as the candidate nodes and rewiring continues in a manner identical to the strategic wiring. The entire scenario occurs according to Algorithm \ref{algo2} and set $\mathcal{E}^{\star \star}$ containing the energies of corresponding LCCs during each rewire operation is obtained. It is obtained from where the candidate links for the rewiring process are chosen. Now, given that total budget $B$ for $r$ rewiring operations, our task is to maximize the energy of the LCC after the rewiring. Hence, the given problem can be formulated into 0/1 Knapsack optimization problem as follows:

\begin{align}
& \max \mbox{  } \sum_{i=1}^p \theta_i x_i \nonumber \\
& \mbox{Subject to } \mbox{  } \sum_{i=1}^p c_ix_i =b \leq B \mbox{ and } C \succ 0 \mbox{ and } x_i \in \{0,1\} \nonumber 
\label{Eq:opt}
\end{align}

where $x$ is the decision variable. For each instance, available budget $b$ is increased by $10 \%$ and theoretical values of Laplacian energy of the LCC is obtained after increment. It is also possible that for the given increment of budget, total available budget is not sufficient for the rewiring. Thus, we get optimal set of $\mathcal{E}^\star$ for the given increment of the budget $b$ using the steps provided in Algorithm \ref{algo1}. Now, for the given $\theta_{i}^{\star} \in \mathcal{E}^\star$, node $i_i \in \mathcal{I}$ is picked and rewiring/addition of the link approach is adopted as in the case of strategic rewiring and node $i_i \in \mathcal{I}$ is connected to $j_i \in \mathcal{J}$. The Laplacian energy set $\mathcal{E}^{\star\star}$ is obtained after each rewire / addition of the link in the underlying network using the steps provided in Algorithm \ref{algo2}.

\begin{algorithm}[!h]
  $\textbf{Function}$ \hspace{2mm} $\textbf{Lenergy}$({b,$\mathcal{C}$,$\mathcal{E}$})\\
  $N$= length($\mathcal{E}$)
	T[N,$b$] $\leftarrow $ 0, K[N,$b$] $\leftarrow $ 0
	
	\For{i $\leftarrow $ 1 to N} 
	{
	\For{c $\leftarrow $ 0 to b} 
	{
	$c_i \leftarrow  \mathcal{C}[i-1]$
	
	$\theta_i \leftarrow  \mathcal{E}[i-1]$
	{
	
	\If{($c_i \leq c)$ and ($\theta_i$ + T[i-1,c-$c_i$] $>$ T[i-1,c])}
    {

		T[i,c] $\leftarrow $ ($\theta_i$ + T[i-1,c-$c_i$])
		
		K[i,c] $\leftarrow $ 1

	}

    \Else
    {
		T[i,c] $\leftarrow $ T[i-1,c]    	
    	 }
    }
    }
    }
    $\mathcal{E}^{\star}$ = [ ]
    
    $b'$= b 
    
    \For{i $\leftarrow $ N downto 1} 
	{
	\If{K[i,$b'$] == 1}
	{
	append $\mathcal{E}[i]$ to $\mathcal{E}^{\star}$	
	
	$b' \leftarrow (b' - \mathcal{C}[i])$
	}
	}
	
	return $\mathcal{E}^{\star}$
\caption{Function to determine $\mathcal{E}^{\star}$}
\label{algo1}
$\textbf{End \hspace{1mm}Function}$
\end{algorithm}

\begin{algorithm}[!h]
\KwInput{$G^{\star} (n,m)$ // one of the largest connected components (LCC) of the fragmented graphs with $n$ nodes and $m$ edges}

$i_{LCC}^{\star}\leftarrow $ $i^{th}$ Node of LCC\\
$j_{LCC}^{\star}\leftarrow $ Neighbor of $i_{LCC}^{\star}$ with maximum degree\\
$N^{\star}\leftarrow$ Length of $\mathcal{E}^{\star}$\\
$A^{\star}$ $/*$ Adjacency matrix of $G^{\star}$ $*/$\\
$j^{\star} \leftarrow $ $j^{th}$ Node belonging to one of the disconnected component \\
$l^{\star} \leftarrow $ Neighbor of $j_{LCC}^{\star}$ with maximum degree with $k_{l^{\star}}\geq 2$\\

\For{i $\leftarrow $ 1 to 10}
{

$b^{\star}=i\ast{0.1}\ast{B}$\\
$\mathcal{E}^{\star}$=$\textbf{Lenergy}$({$b^{\star}$,$\mathcal{C}$,$\mathcal{E}$})\\
\For{j $\leftarrow $ 1 to $N$}
{
\For{k $\leftarrow $ 1 to $N^{\star}$}

{

\If{$\mathcal{E}^{\star}$[k]$==$$\mathcal{E}[j]$}
{
$i_{LCC}^{\star}=\mathcal{I}[k]$\\
$j^{\star}=\mathcal{J}[j]$\\

\If{$m_{G^{\star}}> n$}  
{
$A^{\star}[j_{LCC}^{\star}][l^{\star}]=0$\\
$A^{\star}[l^{\star}][j_{LCC}^{\star}]=0$\\
$A^{\star}[j_{LCC}^{\star}][j^{\star}]=1$\\
$A^{\star}[j^{\star}][j_{LCC}^{\star}]=1$\\
}
\Else
{
$A^{\star}[i_{LCC}^{\star}][j^{\star}]=1$\\
$A^{\star}[i^{\star}][j_{LCC}^{\star}]=1$\\
}
}
}
Update $A^{\star}$, $G^{\star}$\\
Compute $L_{\mathcal{E}}$ of $G^{\star}$\\
Append $L_{\mathcal{E}}$ to $\mathcal{E}^{\star \star}$
}
}
return $G^{r}(n,m)$, $\mathcal{E}^{\star \star}$
\caption{Algorithm to generate connected graph with optimal rewiring under budget constraint} \label{algo2}
\end{algorithm}

\begin{algorithm}[!h]
\KwInput{$\zeta=\{G_1,G2,\dots,G_q\}$; set of $q$ disconnected graphs}
  \KwOutput{Rewired Network $G^{r}$, $C^{r}$, $L^{r}_\mathcal{E}$}
  $LCC \leftarrow$ $G_q$ with maximum number of nodes,$G_q \in \zeta$ \\  
      
	$i_{LCC}^{\star}\leftarrow $ $i^{th}$ Node of LCC with maximum degree\\
    $n \leftarrow$ number of nodes in $G_{q}$ \\
    $m  \leftarrow$ number of edges in $G_{q}$\\
    $u\leftarrow$ node with maximum degree and $u\in G_{q}\neq$ LCC
\While{$|\zeta| > 1$}
{
	 \For{$j \leftarrow  1 \mbox{ to} $ n} 
{
\If{$G_{q}.is\_connected(i_{LCC}^{\star},j) == True$ and $G_{q}.degree(j) >= 2$}
{
$Ne_{j} \leftarrow G_{q}.neighbors(j) $ \\
k $\leftarrow$ Node with maximum degree in $Ne_{j}$ \\
\If{$m > n$}
{
$G_{q}.remove\_edge(k,j)$\\
$G_{q}.add\_edge(K,u)$ \\
$ G^{r} \leftarrow $ LCC \\
$L^{r}_\mathcal{E} \leftarrow evaluate\_energy(G^r)$ \\ 
\Else
{
$G_{q}.add\_edge(k,u)$\\
$ G^{r} \leftarrow $ LCC \\
$L^{r}_\mathcal{E} \leftarrow evaluate\_energy(G^r)$ 
}
}
}
}
}
return $G^{r}, L^{r}_\mathcal{E}$
\caption{Algorithm to generate connected graph under strategic rewiring approach} \label{algo3}
\end{algorithm}
\section{Result and Analysis}
For the simulation, we consider two types of synthetic networks; Random and Power law networks, along with a real-world dataset, the Email-univ network. Initially, all the networks are fragmented into several disconnected components (clusters). Then, the rewiring and edge addition mechanisms (strategic and budget-constrained, as outlined in section \ref{rewire_mech}) are applied to restore the networks. The results and discussions for these networks, focusing on strategic and budget-constrained rewiring, are presented in the following sections.
\subsection{Random network}
In this case, two variants of random networks are considered with the following details:
\begin{enumerate}[(i)]
    \item $n = 500, p = 0.01, m=1277$
    \item $n = 500, p = 0.02, m=2502$
\end{enumerate}
Where $n$ represents the number of nodes, $p$ is the probability of establishing a connection, and $m$ is the total number of edges in the random network. In variant (i), 15 clusters are obtained, while in variant (ii), 6 clusters are formed. Different values of $p$ are selected to determine the varying numbers of disconnected components, allowing for the observation of how the rewiring and edge addition processes work to reconstruct the network using both strategic and budget-constrained approaches.

\begin{figure}[!htb]
\begin{center}
\includegraphics[width=\linewidth,height=4.in]{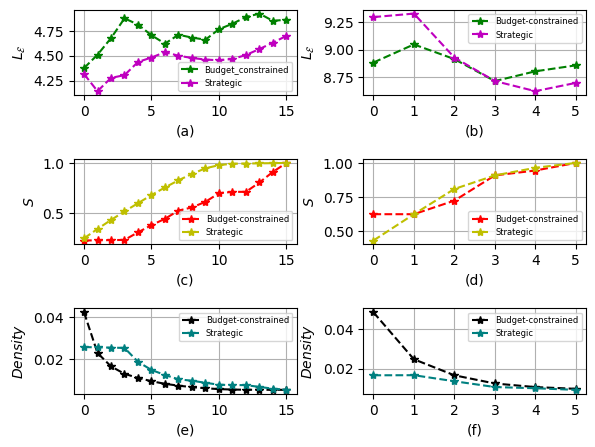}  

\caption{Left panel (a,c,e) represents the plot of parameters $L_{\mathcal{E}}$, $S$ and $\rho$ respectively for the random network having $n=500, p = 0.01$ and right panel (b,d,f) represent the same parameters for  Power Law Network with $n=500, p=0.02$.}
\label{random_0.01_0.02}
\end{center}
\end{figure}

Simulation results  for the parameters $L_{\mathcal{E}}$, $S$ and $\rho$ are shown in Fig. \ref{random_0.01_0.02} (a,c,e) for the setting $p=0.01$ and Figs. \ref{random_0.01_0.02} (b,d,f) for $p=0.02$. In this case, we observe that there is a variation (although small) in the parameter $L_{\mathcal{E}}$ during each step of rewiring for budget and strategic approaches. However, unlike the power-law network, there is no significant drop in $L_{\mathcal{E}}$ as shown in Fig. \ref{random_0.01_0.02} (a). This is due to the difference in the topological structure. In the case of random networks, the average degree of the nodes remains nearly the same and with the $p=0.01$ model and the value of $L_{\mathcal{E}}$ ranges approximately between $4$ and $5$ according to the Eq. \eqref{l_energy_7}, during each step for the rewiring approaches considered as represented in Fig. \ref{random_0.01_0.02}(a). However, in terms of parameter $S$ being concerned, the strategic approach is more promising compared to the budget-constrained approach, as shown in Fig. \ref{random_0.01_0.02}(c). This is because in the subsequent step of rewiring, for the available budget, there may not be the availability of the candidate edge to connect the new component to the existing LCC. In the case of graph density $\rho$, a similar pattern is observed compared to the power-law network; however, due to the topological structure of the random networks, the graph density does not fall to very lower levels during the rewire mechanism, as indicated in Fig. \ref{random_0.01_0.02}(e).

In the case of the random network with $p=0.02$, the variation in $L_{\mathcal{E}}$ is small for both the strategic and budget-constrained approaches. The parameter $L_{\mathcal{E}}$ fluctuates within a range of approximately $8.5$ to $9.5$, as shown in Fig. \ref{random_0.01_0.02}(b), which supports the validity of Eq. \eqref{l_energy_7} for a $d$-regular graph. Additionally, the behavior of the LCC remains consistent during each step of rewiring for the considered approaches, as represented in Fig. \ref{random_0.01_0.02}(d). The difference in graph density $\rho$ becomes negligible after the first rewire, as shown in Fig. \ref{random_0.01_0.02}(f). This suggests that, during the rewiring and edge addition process, the node being connected from the new disconnected component to the existing LCC contributes approximately the same number of edges for both rewiring approaches.

\subsection{Power-law network}
\begin{figure}[!htb]
\begin{center}
\includegraphics[width=1\linewidth,height=4.in]{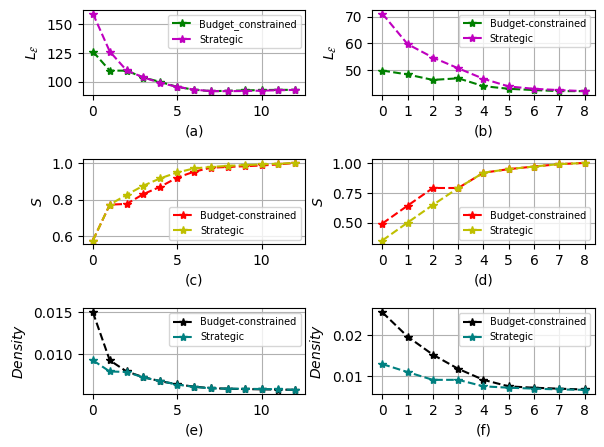}  

\caption{Left panel (a,c,e) represents the plot of parameters $L_{\mathcal{E}}$, $S$ and density $\rho$ respectively for the Power Law Network having $n = 500, \gamma = 2.1$ and right panel (b,d,f) represent the same parameters for  Power Law Network with $n=500, \gamma = 2.7$.}
\label{power_law}
\end{center}
\end{figure}

Here, we consider two variants of the power-law network as follows:
\begin{enumerate}[(i)]
    \item $n = 500, \gamma = 2.1, m=1348$
    \item $n = 500, \gamma = 2.7, m=1865$
\end{enumerate}
where $\gamma$ is the power law exponent, $n$ is the number of nodes and $m$ is number of edges in the network. For the variant (i), there are 12 clusters and for (ii), there are 8 clusters. 
Simulation results  for the parameters $L_{\mathcal{E}}$, $S$ and density $\rho$ are shown in Fig. \ref{power_law}(a,c,e) for the setting $n = 500, \gamma = 2.1$ and Figs. \ref{power_law}(b,d,f) for $n=500, \gamma = 2.7$. For the power-law network with $\gamma=2.1$, we get twelve connected components, therefore total twelve rewiring/addition of edges take place depending on nodes and edges in LCC after each step. It is observed from Fig. \ref{power_law}(a) that the value of the parameter $L_{\mathcal{E}}$ is approximately $159$ for strategic rewiring and $126$ for the budget-constrained rewiring approach. After the second rewiring/addition of edge to the existing LCC, $L_{\mathcal{E}}$ is nearly identical for both strategic and budget-constrained rewiring approaches as shown in Fig. \ref{power_law}(a). In Fig. \ref{power_law}(c), the effect of subsequent rewiring and edge addition on the size of the LCC is shown, where the value of $S$ continues to increase. After the final step, $S=1$, indicating that all the nodes in the network are now part of the LCC. When an edge is rewired, no additional edge is added to the existing LCC. However, depending on the number of nodes and edges in the current LCC, a new edge may be introduced to connect the node with the maximum degree in the LCC to the node with the highest degree in the new component being added. During each step of the rewiring and edge addition process, the number of nodes in the LCC increases, but the number of edges does not increase significantly. As a result, the graph density $\rho$ continues to decrease, as shown in Fig. \ref{power_law}(e). After the complete rewiring for the $\gamma = 2.1$ model, the LCC consists of $m=707$ edges, which implies that the network is restored by adding only a few edges and rewiring existing edges for both the strategic and budget-constrained approaches. 

For the $\gamma=2.7$ model, initially (up to step 4), there is variation in the parameter $L_{\mathcal{E}}$ as represented in Fig. \ref{power_law}(b). It occurs due to the availability of candidate edge for rewire/addition operation for strategic and budget constrained approaches. For the budget-constrained approach, with each increment of the budget, a new component is added to the existing LCC and the parameter $L_{\mathcal{E}}$ is updated as shown in the last row of the Table \ref{table1}. Thus, for each rewiring action, the parameter $S$ keeps on increasing and $S=1$ in the last step showing that all the nodes of the network are connected to the LCC. Similarly, graph density $\rho$ varies up to step 4 for strategic and budget constrained approaches and keeps on decreasing in each step. This is primarily due to the larger number of nodes in the LCC compared to the number of edges. For the $\gamma = 2.7$ model, after the entire network is restored (rewired), the total number of edges is $m=831$. Thus, we observe that the network is reconstructed with fewer than 50\% of the edges, making the rewiring process highly economical, especially in situations where the addition of edges becomes more costly.
\begin{table}[]
\centering
\caption{First column represents the available budget (in units) after each increment of $10 \%$. Columns 0-8 represent the updation in the parameter $L_{\mathcal{E}}$ by the addition of the new component to the existing LCC after subsequent rewire/addition operation. }
\label{table1}
\resizebox{\columnwidth}{!}{%
\begin{tabular}{|c|c|c|c|c|c|c|c|c|c|}
\hline
\textbf{Budget} & \textbf{0} & \textbf{1} & \textbf{2} & \textbf{3} & \textbf{4} & \textbf{5} & \textbf{6} & \textbf{7} & \textbf{8} \\ \hline
\textbf{112} & \textbf{} & \textbf{} & \textbf{} & \textbf{} & \textbf{} & \textbf{} & \textbf{} & \textbf{} & \textbf{} \\ \hline
\textbf{224} & \textbf{71.08} & \textbf{} & \textbf{} & \textbf{} & \textbf{} & \textbf{} & \textbf{} & \textbf{} & \textbf{} \\ \hline
\textbf{336} & \textbf{59.74} & \textbf{71.08} & \textbf{} & \textbf{} & \textbf{} & \textbf{} & \textbf{} & \textbf{} & \textbf{} \\ \hline
\textbf{449} & \textbf{54.77} & \textbf{59.74} & \textbf{71.08} & \textbf{} & \textbf{} & \textbf{} & \textbf{} & \textbf{} & \textbf{} \\ \hline
\textbf{561} & \textbf{50.78} & \textbf{54.77} & \textbf{59.74} & \textbf{71.08} & \textbf{} & \textbf{} & \textbf{} & \textbf{} & \textbf{} \\ \hline
\textbf{673} & \textbf{46.86} & \textbf{50.78} & \textbf{54.77} & \textbf{59.74} & \textbf{71.08} & \textbf{} & \textbf{} & \textbf{} & \textbf{} \\ \hline
\textbf{786} & \textbf{44.02} & \textbf{46.86} & \textbf{50.78} & \textbf{54.77} & \textbf{59.74} & \textbf{71.08} & \textbf{} & \textbf{} & \textbf{} \\ \hline
\textbf{898} & \textbf{43.15} & \textbf{44.02} & \textbf{46.86} & \textbf{50.78} & \textbf{54.77} & \textbf{59.74} & \textbf{71.08} & \textbf{} & \textbf{} \\ \hline
\textbf{1010} & \textbf{42.67} & \textbf{43.15} & \textbf{44.02} & \textbf{46.86} & \textbf{50.78} & \textbf{54.77} & \textbf{59.74} & \textbf{71.08} & \textbf{} \\ \hline
\textbf{1123} & \textbf{42.28} & \textbf{42.67} & \textbf{43.15} & \textbf{44.02} & \textbf{46.86} & \textbf{50.78} & \textbf{54.77} & \textbf{59.74} & \textbf{71.08} \\ \hline
\end{tabular}%
}
\end{table}

\subsection{Email-univ network}
The real-world dataset used in this study consists of email communications within a university located in the southern region of Catalonia, Spain. In this dataset, the university users are represented as nodes, with a link between two nodes indicating that they have communicated at least once via email. The detailed description about the dataset is provided in Table \ref{tab:datasetinfo}. Before the attacks, the network contained $1133$ nodes and $5154$ edges. However, after the attack, the network was divided into nine disconnected components, requiring at least nine rewiring steps to reconnect the network. After the rewiring process (both strategic and budget-constrained), the network is restored with approximately 43.45\% (2234) of the original edges.

\begin{table}[!htb]
    \centering
     \caption{Dataset Information}
    \begin{tabular}{|p{4.0 cm}|p{1.5 cm}|}
        \hline
       Nodes ($n$)  & $1133$ \\\hline
       Edges ($m$)  & $5451$ \\\hline
       Average degree  &	$9$ \\\hline
       Average shortest path length & $3.606$ \\ \hline
    \end{tabular}
   
    \label{tab:datasetinfo}
\end{table}

\begin{figure*}[!htb]
\begin{center}
\includegraphics[width=0.95\linewidth,height=2.in]{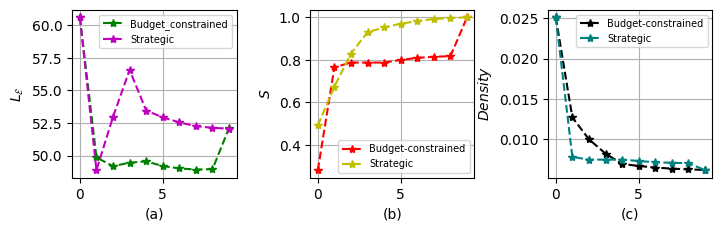}  
\caption{The plot of parameters $L_{\mathcal{E}}$, $S$ and density $\rho$ respectively for the email university network having $n = 1133, \mbox{ } m = 5154$.}
\label{email_real.png}
\end{center}
\end{figure*}

At the initial stage of rewiring, the values of the parameter $L_{\mathcal{E}}$ are almost same for both the rewiring methods. However, with the increase in the number of rewiring, $L_{\mathcal{E}}$ increases up to third rewire and then decreases and $L_{\mathcal{E}}\approx 52$ at the last rewire for the strategic approach. Whereas, for the budget-constrained approach, $L_{\mathcal{E}}$ lies between $40$ and $50$ in the last rewire $L_{\mathcal{E}}\approx 52$ as shown in Fig. \ref{email_real.png}(a). For the robustness index parameter $S$, both the rewiring approaches show promising results by getting $S\approx 0.8$ after only third rewire. For the strategic rewiring approach, the value of the parameter $S$ increases with each rewire, reaching $S=1$ after the final rewire. However, for the budget-constrained approach, after the third rewire, $S$ remains unchanged until the last rewire. This happens because, as the budget increases, the available links do not contribute to connecting nodes from the disconnected components to the existing largest connected component, as shown in Fig. \ref{email_real.png}(b). Regarding density $\rho$, for the strategic approach, it drops significantly after the first rewire and then remains constant for the subsequent rewiring steps. This is due to the addition of more nodes with relatively few edges to the existing LCC. In contrast, for the budget-constrained approach, density decreases gradually, as shown in Fig. \ref{email_real.png} (c).

\section{Conclusions and Future Works}
In the presented research work, a method is proposed to compute the edge weights based on the reputation between the pair of nodes. These edge weights are quantified as the cost of construction of the links. Further, the research works emphasize mainly the restoration of the networks (after the attacks) under rewiring (strategic and budget-constrained) mechanisms. While the rewiring mechanism, it is taken care that the candidate link chosen from the existing largest connected component does not result in further disconnected components. In the worst case, if it happens then an additional link is introduced. However, this scenario happens very less number of times. The proposed rewiring mechanisms enable the restoration of the fragmented network by utilizing nearly $40-50\%$ of the total links. Thus, the proposed methods avoid the additional overheads of introducing additional links to restore the network. During the restoration of the networks, robustness is measured in-terms of robustness index $S$ and Laplacian energy $L_{\mathcal{E}}$. Although, $S$ measures the size of LCC, however it does not quantify how fragile is the network after restoration. Therefore, the parameters  $L_{\mathcal{E}}$ and edge density $\rho$ are introduced to evaluate the sparsity of the restored networks. From the simulation result of the considered synthetic and real-world dataset networks, it is evident that $L_{\mathcal{E}}$ and $\rho$ are positively correlated. Hence, it is natural that while evaluating the robustness, along with the size of the largest connected components, the sparsity of the underlying network should also be considered.

The learning concept can be used to predict future network robustness by considering the current scenario. Hence, the work can be extended by considering machine learning approach-based rewiring operations. Apart from this, the work can be used for robustness analysis for multilayer network. The proposed approaches can be used for analyzing the performance of other real-world applications for generalizing the performance of complex networks.


\ifCLASSOPTIONcaptionsoff
  \newpage
\fi


\bibliographystyle{IEEEtran}
\bibliography{ref}

\end{document}